\documentclass[preprint,5p,twocolumn,authoryear]{elsarticle}

\usepackage{epsfig}
\usepackage{natbib}

\journal{Advances in Space Research}

\begin{document}

\begin{frontmatter}

\title{Study of hot flow anomalies using Cluster multi-spacecraft measurements}
\author[lpc2e]{G.~Facsk\'{o}\corref{gfacsko}}
\author[lpc2e]{J.~G.~Trotignon}
\author[cesr]{I.~Dandouras}
\author[ic]{E.~A.~Lucek}
\author[lindau]{P.~W.~Daly}
\ead{gfacsko@cnrs-orleans.fr}
\cortext[gfacsko]{Corresponding author}

\address[lpc2e]{LPC2E/CNRS, 3A, Avenue de la Recherche Scientifique, 45071 Orl\'eans cedex 2, France}
\address[cesr]{CESR/CNRS, 9, Avenue du Colonel Roche, 31028 Toulouse cedex 4, France}
\address[ic]{The Blackett Laboratory, Imperial College London, Prince Consort Road, London SW7 2BW, UK}
\address[lindau]{Max-Planck-Institut f\"ur Aeronomie, Max-Planck-Str. 2, 37191 Katlenburg-Lindau, Germany}

\begin{abstract}
Hot flow anomalies (HFAs) were first discovered in the early 1980s at the bow shock of the Earth. In the 1990s these features were studied, observed and simulated very intensively and many new missions (Cluster, THEMIS, Cassini, and Venus Express) focused the attention to this phenomenon again. Many basic features and the HFA formation mechanism were clarified observationally and using hybrid simulation techniques.  We described previous observational, theoretical and simulation results in the research field of HFAs. We introduced HFA observations performed at the Earth, Mars, Venus and Saturn in this paper. We share different observation results of space mission to give an overview to the reader.

Cluster multispacecraft measurements gave us more observed HFA events and finer, more sophisticated methods to understand them better. In this study HFAs were studied using observations of the Cluster magnetometer and the Cluster plasma detector aboard the four Cluster spacecraft. Energetic particle measurements (28.2-68.9\,keV) were also used to detect and select HFAs. We studied several specific features of tangential discontinuities generating HFAs on the basis of Cluster measurements in the period February-April 2003, December 2005-April 2006 and January-April, 2007, when the separation of spacecraft was large and the Cluster fleet reached the bow shock. We have confirmed the condition for forming HFAs, that the solar wind speed is higher than the average. This condition was also confirmed by simultaneous ACE magnetic field and solar wind plasma observations at the L1 point 1.4 million km upstream of the Earth. The measured and calculated features of HFA events were compared with the results of different previous hybrid simulations. During the whole spring season of 2003, the solar wind speed was higher than the average. Here we checked whether the higher solar wind speed is a real condition of HFA formation also in 2006 and 2007. 

At the end we gave an outlook and suggested several desirable direction of the further research of HFAs using the measurements of Cluster, THEMIS, incoming Cross Scale and other space missions. 
\end{abstract}


\end{frontmatter}

\section{Introduction}
\label{sec:intro}

Hot flow anomalies (HFAs) were discovered in the early 1980s using AMPTE and ISEE measurements \citep{schwartz85,thomsen86:_hot}. In the 1980s and the early 1990s they were studied very intensively. After 2000 researchers left this topic. In the last few years, however HFAs became being studied intensively again thanks for the Cluster, THEMIS, Mars Global Surveyor (MGS), Voyager, Venus Express (VEX) and Cassini missions. The purpose of this review is to present some of these results and give an overview to the reader about hot flow anomalies. 

\subsection{Observations}
\label{sec:observation}

\begin{figure}[htb]
\centering
\epsfig{file=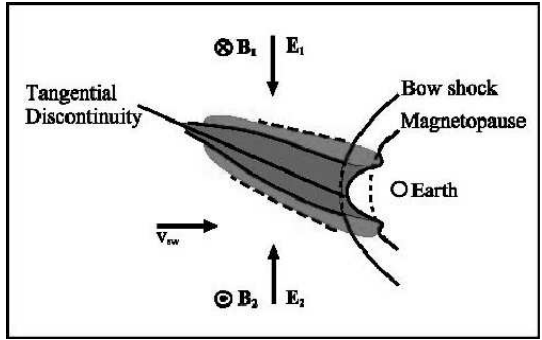,width=200pt}  
\caption{The relative position of the tangentional discontinuity (TD) and the bow shock (BS) of the Earth during HFA formation; furthermore the direction of the convective electric field. This HFA was observed by THEMIS spacecraft. \citep{eastwood08:_themis_obser_of_hot_flow_anomal}}
\label{fig:hfa1}
\end{figure}

\begin{figure}[b]
\centering
\epsfig{file=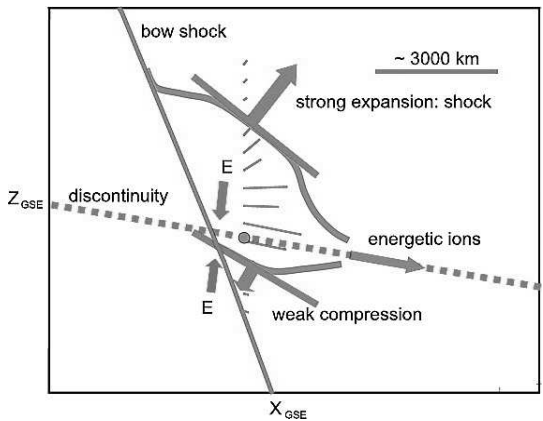,width=200pt}  
\caption{The environment of the interaction zone: the cavity and its propagation observed by Cluster. \citep{lucek04:_clust}}
\label{fig:hfa2}
\end{figure}

The fundamental properties of HFAs were clarified by \citet{schwartz95:_hot_flow_anomal_near_earth_bow_shock}. A tangentional discontinuity (TD) interacts to the bow shock (BS) and a tensious, hot bubble forms (See Fig.~\ref{fig:hfa1}). The original name of HFAs was hot diamagnetic cavity (HDC) however they are also known as active currents sheets (AC) because the HFA extends alongside the discontinuity \citep[See also Fig.~\ref{fig:hfa1}]{eastwood08:_themis_obser_of_hot_flow_anomal}. HFAs have significant influence on the magnetopause (MP) \citep{sibeck99:_compr, sibeck00:_magnet_motion_driven_by_inter, sibeck01:_solar, eastwood08:_themis_obser_of_hot_flow_anomal}. The outer surface of the expanding cavity behaves as fast mode shock and the inner surface of the expanding region shows the features of a tangentional discontinuity \citep[See also Fig.~\ref{fig:hfa2}]{schwartz95:_hot_flow_anomal_near_earth_bow_shock,paschmann88:_three,lucek04:_clust}. The electron distribution is an isotropic Maxwell distribution \citep{thomsen86:_hot,schwartz95:_hot_flow_anomal_near_earth_bow_shock} whose state is reached by fire hose instability \citep{eastwood08:_themis_obser_of_hot_flow_anomal}. The proton distribution is Maxwellian \citep{thomsen86:_hot,schwartz95:_hot_flow_anomal_near_earth_bow_shock}. The wave properties of the HFAs were studied by \citet{tjulin08:_wave_activ_insid_hot_flow_anomal_cavit} in different stage of their evolution using k-filtering techniques and multispacecraft measurements from the Cluster fleet. The hot plasma of the expanding cavity can be considered turbulent and shows non-linear properties \citep{kovacs09:_turbul_behav_of_hot_flow}. The typical change angle of the magnetic field through the discontinuity is $70^o$ \citep{schwartz95:_hot_flow_anomal_near_earth_bow_shock,safrankova02,facsko09:_global_study_of_hot_flow,facsko08:_statis_study_of_hot_flow}. HFA propagate into the magnetosheath (MS) \citep{safrankova02} and along the TD \citep{eastwood08:_themis_obser_of_hot_flow_anomal}. The active current sheet expands farther in the upstream region \citep{safrankova02}. The system is not in pressure balance so the events expand \citep{thomsen86:_hot,lucek04:_clust}. Sometimes double HFAs are observed \citep{safrankova00:_magnet}. Suprathermal particles may be present \citep{lucek04:_clust,kecskemety06:_distr_rapid_clust} but energetic particle events are not observed in all cases. When they are observed then the flux increase starts before the magnetic signature and ends after the magnetic perturbations \citep{kecskemety06:_distr_rapid_clust}. The beam generated by the interaction of the discontinuity and the bow shock is usually parallel to the TD surface \citep{lucek04:_clust}. 

\subsubsection{Statistical studies}
\label{sec:stat}

Several statistical studies of HFAs were carried out before the Cluster mission. These studies were based on 9 to 20 events because it was very difficult to observe these dramatic perturbations of the solar wind plasma that can be clearly identified with the appropriate instruments. HFAs were considered as rare events after their discovery; later an average of 3 events was supposed to occur per day \citep{fuselier87:_fast,schwartz00:_condit}. This opinion did not have to be revised after the launch of Cluster, which revealed many HFA events \citep{kecskemety06:_distr_rapid_clust,facsko08:_statis_study_of_hot_flow,facsko09:_global_study_of_hot_flow,facsko09:_clust_hot_flow_anomal_obser}. Now we know that HFAs appear very frequently and when several conditions are satisfied, like the large angle between the normal of the TD and the Sun-Earth direction \citep{schwartz00:_condit,safrankova02} and a high solar wind speed \citep{safrankova00:_magnet,facsko08:_statis_study_of_hot_flow,facsko09:_global_study_of_hot_flow,facsko09:_clust_hot_flow_anomal_obser}. Besides of the higher solar wind speed the TD must sweep slowly past the surface of the BS \citep{schwartz00:_condit,facsko09:_clust_hot_flow_anomal_obser,facsko09:_global_study_of_hot_flow}. This feature is almost always coupled to high solar wind speed beams \citep[downstream:][]{safrankova00:_magnet, facsko08:_statis_study_of_hot_flow} \citep[upstream:][]{facsko08:_statis_study_of_hot_flow,facsko09:_clust_hot_flow_anomal_obser,facsko09:_global_study_of_hot_flow}. 

HFAs propagate and extend both in the magnetosheath and in the upstream region. The statistical study by \citet{safrankova02} suggests that HFAs impact $10\,R_{\mathrm{g}}$ into the magnetosheath and extend approximately $25\,R_{\mathrm{g}}$ in the solar wind, where $R_{\mathrm{g}}$ is the gyroradius. \citet{facsko09:_global_study_of_hot_flow} discovered HFAs at a much larger distance from the bow shock. These HFAs are triggered by tangential but not rotational discontinuities \citep{safrankova02}. Double HFAs can be observed both in the solar wind and the magnetosheath \citep{safrankova02}. The convective electric field ($\mathbf{E}=-\mathbf{v}\times\mathbf{B}$) is very important in the development of HFAs: the field focuses particles on the TD and the discontinuity leads them back to the perpendicular bow shock where they are accelerated \citep{thomsen93:_obser_test_hot_flow_abnom}. If the electric field is symmetric then the cavity is also symmetric. If the field is asymmetric then an asymmetric cavity is the result. If the electric field is not present at any side of the TD then HFAs cannot develop \citep{thomas91:_hybrid,thomsen93:_obser_test_hot_flow_abnom}. The associated energetic particles were studied by \citet{louarn03:_obser_of_energ_time_disper} using Cluster RAPID \citep[Research with Adaptive Particle Imaging Detectors;][]{wilken01:_first_resul_from_rapid_imagin} measurements in the magnetosheath. Particle acceleration was explained by first order Fermi acceleration. Nowadays the quasi-perpendicular bow shock is considered as the prime source of the energy of the energized particles \citep{lucek04:_clust,omidi07:_format,nemeth08:_partic_accel_at_inter_of}. 

\subsubsection{Extra-Terrestrial HFAs}
\label{sec:et}

\begin{figure}[ht]
\centering
\epsfig{file=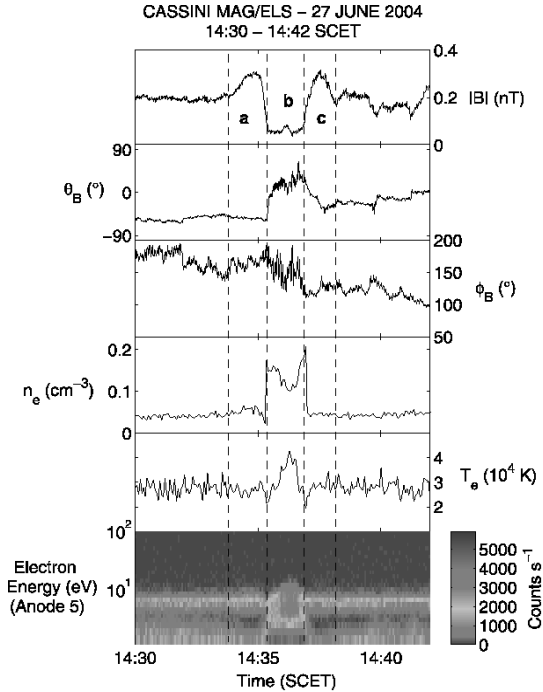,width=200pt}  
\caption{A HFA analogy in the Kronian system. The top three panels show the magnitude and direction of the magnetic field in spherical polar coordinates. The fourth and fifth panels are the electron number density and temperature respectively and the sixth (bottom) panel is a time-energy spectrogram of electron count. \citep[Fig.~1]{masters08:_cassin}}
\label{fig:et}
\end{figure}

Fortunately these hot diamagnetic cavities do not occur only at the terrestrial bow shock. HFA would form wherever there is an appropriate interaction between an interplanetary tangentional discontinuity and a collisionless shock as \citet{lucek04:_clust} proposed. Using measurements from MGS \citep[Mars Global Surveyor;][]{albee98:_mars_global_survey_mission} the Martian analogies were discovered by \citet{oeieroset01:_hot_martian} at the flank of Martian bow shock. Two events were observed and analyzed which looked like HFAs in the magnetic field and electron data, however the dramatic ion heating, and bulk flow deflection, typical of terrestrial HFAs were unable to be shown. This result led \citet[submitted on November 4, 2006]{facsko08:_statis_study_of_hot_flow} to suggest looking for HFAs near other planets, for example at Saturn using Cassini MAG \citep[MAGnetometer;][]{dougherty04:_cassin_magnet_field_inves} and CAPS \citep[Cassini Plasma Spectrometer;][]{young04:_cassin_plasm_spect_inves} measurements. Finally, some events were discovered by \citet{masters08:_cassin} (See: Fig.~\ref{fig:et}). Due to pointing constraints the CAPS instrument was unable to measure the solar wind ion distribution so the plasma velocity had to be measured by the electron instrument. The electron density was found to increase inside the cavity instead of dropping. The geometry of the Kronian HFAs differs from terrestrial ones because the different features of the bow shock of Saturn. Their size was measured larger than at Earth in units of km but in units of gyro radii their size may be comparable \citep{masters08:_cassin}. Both must be generated by the same mechanism, however, because the calculated convective electric field points and focuses the ions to the discontinuity on both sides of the TD. During the Venus flyby, MESSENGER observed an active current sheet which could be HFA \citep{slavin09:_messen_and_venus_expres_obser}. Voyager-1, which approached and crossed the termination shock observed an event that can be considered as hot flow anomaly \citep{burgess08:_hybrid_simul_of_inter_of}. So HFAs do not occur in the Solar System only but also in the interstellar space. 

These events observed at Mars (by MGS), at Saturn (by Cassini), at Venus (by Messenger), and at the termination shock (by Voyager-1), were all HFA-like, however none have been shown to have the same dramatic ion heating and deflection of solar wind flow as terrestrial HFAs yet. This means that HFAs have not been confirmed at another bow shock yet, although the Kronian events satisfy the conditions for HFA formation \citep{schwartz00:_condit,facsko08:_statis_study_of_hot_flow,facsko09:_global_study_of_hot_flow,facsko09:_clust_hot_flow_anomal_obser} and are anticipated to be examples of the same phenomenon. 

\subsection{Theory and simulations}
\label{sec:theory}

A HFA is a complex non-linear structure and so its theory is based on observationally confirmed numerical simulations. The ions play an important role in the processes of the events so the simulations can be test particle, full particle and hybrid simulations. Test particle simulations were performed by \citet{burgess89}. The position of the tangentional discontinuity and a flat bow shock were fixed. After setting the configuration of the magnetic field charged particles were injected into the simulation space. The simulation describes the trajectories of the ions in the environment of the discontinuities. 

Hybrid simulations - where electrons are considered as fluid - can be more useful for studying HFAs \citep{thomas88:_evolut,thomas89:_three,thomas91:_hybrid,lin97:_gener,lin02:_global,lin03:_global,onsager91:_inter_reflec,omidi07:_format}. In the nineties supercomputers were needed even for 1D \citep{thomas91:_hybrid} or 2D hybrid simulations but nowadays a personal computer can do the job. Increasingly sophisticated codes help to understand detailed processes associated with HFAs. The role of electric field was first discussed  by \citet{thomsen88:_origin_of_hot_diamag_cavit, thomsen93:_obser_test_hot_flow_abnom} and \citet{thomas91:_hybrid}.  \citet{lin97:_gener}'s study clarified whether rotational or tangential discontinuity interacts with the bow shock. Particle acceleration and wave-particle interactions were also modelled by hybrid simulations \citep{onsager91:_inter_reflec,thomas91:_hybrid}. 

\begin{figure}[ht]
\centering
\epsfig{file=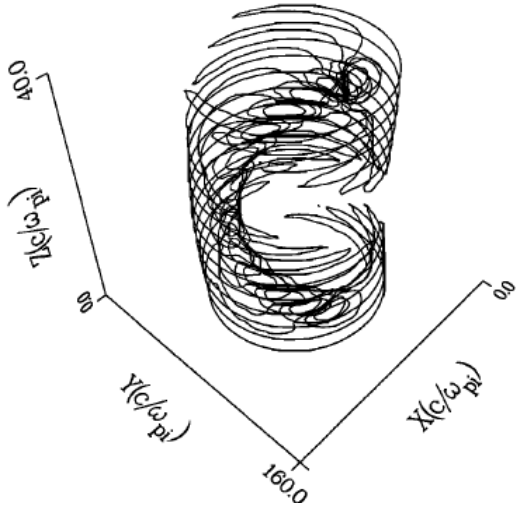,width=200pt}  
\caption{3D hybrid simulation shows the equal density surface of the diamagnetic cavity formed by low frequency Alfv\'en waves \citep{thomas89:_three}.}
\label{fig:alfven}
\end{figure}

The cavity of a HFA is considered as turbulent region \citep[Fig.~\ref{fig:alfven}]{thomas88:_evolut} and simulations are indeed confirmed observationally \citep{kovacs09:_turbul_behav_of_hot_flow}. 2D hybrid simulations suggest that HFAs are heated by low frequency waves and by electron-ion instability. The interaction regions eject a beam whose energy is transferred to Alfv\'en waves that propagate perpendicularly to the direction of the beam. The cavity is not formed by the thermal pressure: 1/3 of the energy is for heating and the rest of it is for waves. The waves create the cavity in the solar wind and the magnetosheath. The size of the cavity is proportional to the energy of the beam. The velocity distribution of electrons is isotropic inside the cavity and the ion distribution seems to be Maxwellian. The finite size of the beam is the reason of the development of the cavity \citep{thomas88:_evolut}. 3D hybrid simulations confirmed the previous result. The Alfv\'en waves remained perpendicular to the beam. The instability also plays an important role. A density wave starts from the center \citep{thomas89:_three}. 

\citet{lin02:_global} studied numerically the interaction for different orientations and configurations of the tangentional discontinuity at the bow shock. She noted that the magnetopause had a bulge as \citet{sibeck01:_solar} and \citet{eastwood08:_themis_obser_of_hot_flow_anomal} observed. The angle ($\gamma$) between the Sun direction  is large and HFA cannot form if this cone angle is less than $43^o$ or larger than $83^o$ in her simulation. The angle between the directions of the magnetic field vectors on the two sides of the discontinuity (so called change angle, $\Delta\Phi$) was large approximately $70^o$ as it was observed by \citet{schwartz95:_hot_flow_anomal_near_earth_bow_shock,safrankova02,facsko08:_statis_study_of_hot_flow,facsko09:_global_study_of_hot_flow}. She also found the ion-beam instability to be the source of the energy. The electric field points to the discontinuity and depends on the change angle. If the change angle is larger then the size of the cavity also increases. This size increases with solar wind speed and also depends on $\gamma$. She found a maximum size when $\gamma=80^o$. 

\begin{figure}[ht]
\centering
\epsfig{file=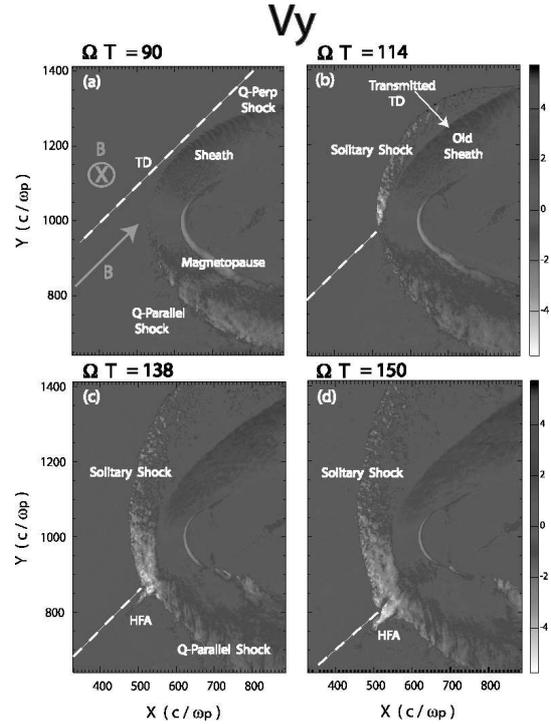,width=200pt}  
\caption{Different phases in the generation of a HFA in the hybrid simulations of \citet{omidi07:_format}. The beam and the importance of the boundary region of the quasi-parallel and perpendicular region are evident.}
\label{fig:omidi}
\end{figure}

Finally it seems that HFAs appear at the boundary surface of the quasi-parallel and quasi-perpendicular region of bow shocks. At the quasi-perpendicular bow shock particles are accelerated and the escaping ions are trapped and focused by the tangentional discontinuity. A fraction of them are accelerated many times. This process ends when the TD reaches the quasi-parallel region \citep[Fig.~\ref{fig:omidi}]{omidi07:_format}. A beam is then ejected and its energy is dissipated by electromagnetic ion-beam instabilities if the size of the beam is limited. Alfv\'en waves generate a cavity and the rest of the energy heats the plasma \citep{thomas88:_evolut,thomas89:_three}. The interaction of the TD and the bow shock led us to discover a new type of the shock called solitary shock, where the TD changes the behaviour of the shock and magnetosheath \citep{omidi07:_format}.

\section{Cluster multi-spacecraft measurements}
\label{sec:multi}

Only few HFA studies were carried using multispacecraft measurements. \citet{safrankova00:_magnet} used the observations of INTERBALL-1 and its subsatellite, MAGION-4. These measurements however cannot suffer from the fact that both satellites had different instruments. The first real multispacecraft HFA observations were made by \citet{lucek04:_clust}, using Cluster measurements. The advantage of four spacecraft was used in their work. The normal of the TD was calculated from the relative observation time and positions of the Cluster fleet. They could determine the transition speed of the intersection line of the TD and the bow shock ($\sim$110\,km/s) which was according to the previous observations and theory \citep{schwartz00:_condit}. The different spacecraft really observed the motional electric field calculated from the measured magnetic field and solar wind velocity vectors. The compression and propagating direction and speed of the cavity boundaries were also observed by the Cluster fleet. The CIS \citep[Cluster Ion Spectrometry;][]{reme01:_first_multis_ion_measur_in} measured the velocity distributions of the HFAs and they observed two particle populations coupled to the solar wind and the ions streaming away from the terrestrial bow shock. Using these observations they determined the age of the observed HFAs.

\citet{kecskemety06:_distr_rapid_clust} concentrated on energized particles measured by RAPID, \citet{kovacs09:_turbul_behav_of_hot_flow} studied the turbulence inside the cavity. \citet{tjulin08:_wave_activ_insid_hot_flow_anomal_cavit} used Cluster as a wave telescope to investigate the wave properties of HFAs. Actually the Cluster mission \citep{escoubet97:_clust_scien_and_mission_overv} seems to be good platform for wave studies because four spacecraft measures simultaneously from different points. This enables to identify wave vectors directly in 3D. In \citet{tjulin08:_wave_activ_insid_hot_flow_anomal_cavit}'s study k-filtering technique was used \citep{pincon91:_local_charac_of_homog_turbul}. The events studied in \citet{tjulin08:_wave_activ_insid_hot_flow_anomal_cavit} were also studied previously in \citet{lucek04:_clust} particularly. One of them was considered young and the other event was a more developed HFA. The k-filtering technique observed two particle populations in the less developed event and only one population in the older HFA. The first population was associated to the solar wind and the second particle population was specularly reflected at the bow shock. The older HFA has only one, hot and deflected particle population. Naturally the presence of the two populations was unstable so they developed and united into one population heating the plasma in the cavity. Only special wavelengths were observed so using the CIS solar wind measurements the size of the HFA could be estimated. It was only 1400\,km however this size does not contradict the simulations \citep{lin02:_global} and the other size estimations \citep{facsko09:_global_study_of_hot_flow} of developed HFAs because this event was quite young. The plasma parameters may have influence for the periodicity so the gyroradius was calculated using the mean magnetic field inside the HFA measured by FGM \citep[Fluxgate Magnetometer;][]{balogh01:_clust_magnet_field_inves}. The result (2200\,km) is the maximum that can be accepted as scaling factor of the periodicy. This might mean that the HFA size and the plasma parameters together determine and filter the wave length inside the cavity.

\citet{eastwood08:_themis_obser_of_hot_flow_anomal} performed the first multispacecraft HFA study based on THEMIS measurements, in which HFAs were observed in the magnetosheath and in the upstream region simultaneously. Cluster has the advantage of having fully identical spacecraft, with instruments covering a wide range of conditions. The analyzing and processing of the Cluster magnetic, plasma and energized particle observations led to the largest statistical study of HFAs \citep{facsko08:_statis_study_of_hot_flow,facsko09:_global_study_of_hot_flow,facsko09:_clust_hot_flow_anomal_obser}. 

\subsection{Selection of HFA events}
\label{sec:selection}

\begin{figure}[ht]
\centering
\epsfig{file=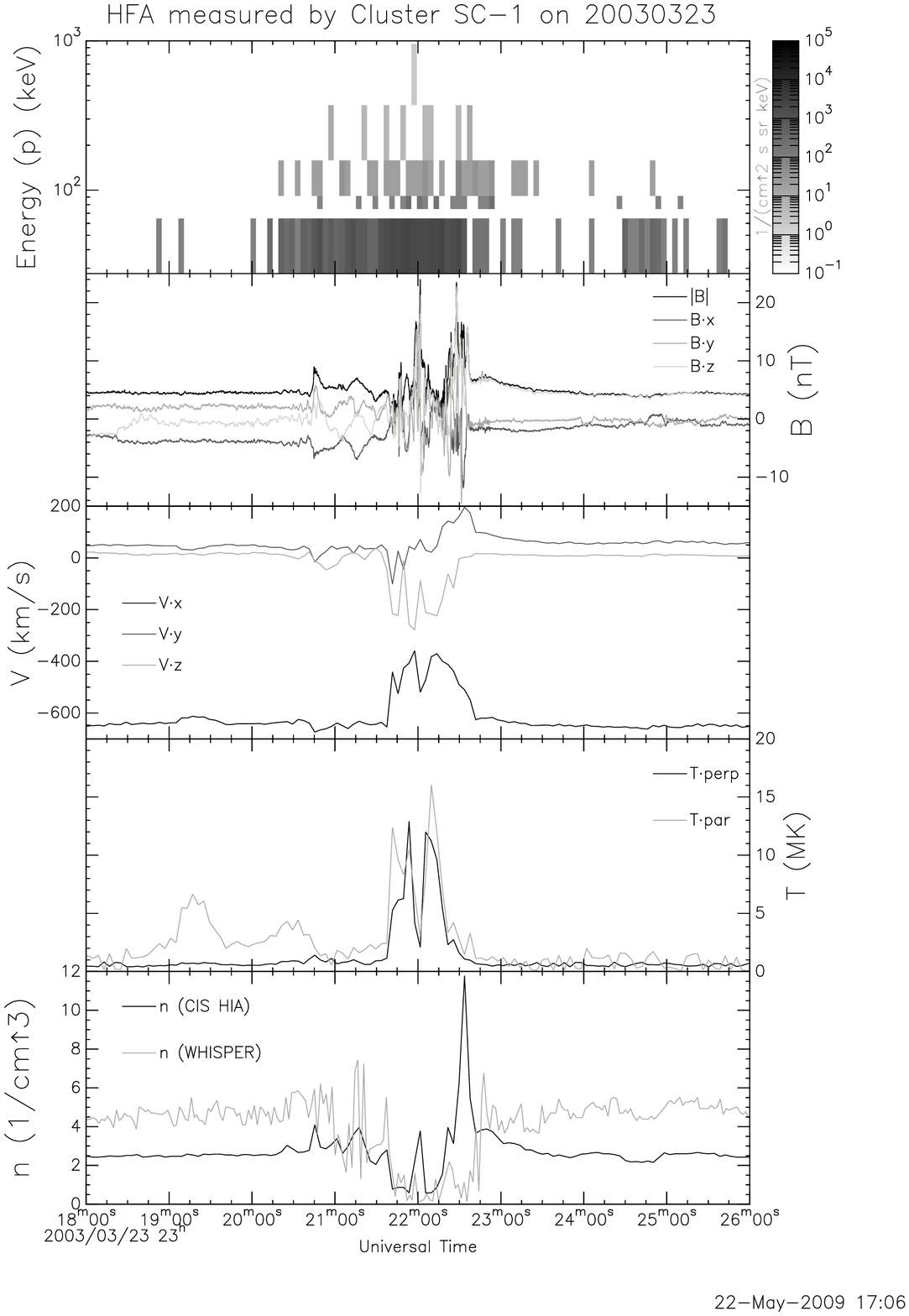,width=250pt}  
\caption{A hot flow anomaly event occurred at 23:22 UT on March 23, 2003 as measured by Cluster 1 RAPID, FGM, CIS HIA and WHISPER \citep[Waves of HIgh frequency and Sounder for Probing of Electron density by Relaxation;][]{decreau01:_early_resul_from_whisp_instr_clust} instruments. Top panel: energetic particle fluxes (in $p/\left({cm}^{3}\,s\,sr\,keV\right)$). Second panel: magnetic field components (in nT). Third panel: solar wind speed components  (in GSE, km/s). Fourth panel: solar wind temperature parallel and perpendicular to $\mathbf{B}$ (MK). Bottom panel: plasma CIS HIA ion and WHISPER electron density ($cm^{-3}$).}
\label{fig:hfasc}
\end{figure}

Based on the above-mentioned observational studies \citep[See also Fig.~\ref{fig:hfasc}]{facsko08:_statis_study_of_hot_flow,facsko09:_global_study_of_hot_flow,facsko09:_clust_hot_flow_anomal_obser,schwartz85,thomsen86:_hot,thomsen93:_obser_test_hot_flow_abnom,sibeck99:_compr,sibeck02:_wind} HFAs can be selected based on the following criteria:
\begin{enumerate}
\item The rim of the magnetic cavity must be visible as a sudden increase in the magnetic field magnitude. Inside the cavity the magnetic field value drops and its direction turns around.
\item The solar wind slows down and its direction always turns away from the Sun-Earth direction.
\item The solar wind temperature increases and its value reaches up to several ten million degrees inside the cavity. 
\item The solar wind particle density also increases on the rim of the cavity and drops inside the HFA. 
\item The tangential discontinuity must appear in the upstream magnetic field.
\end{enumerate}
Often but not always energized particles are present mostly in the range of 28.2-68.9\,keV however sometimes in the range of 68.9-96.1\,keV and 96.1-169.3\,keV. Particle events start before the magnetic signatures appear and end after them \citep{kecskemety06:_distr_rapid_clust}. The convective electric field must point towards the discontinuity on both sides of the TD \citep{thomsen93:_obser_test_hot_flow_abnom}. 

\begin{figure}[ht]
\centering
\epsfig{file=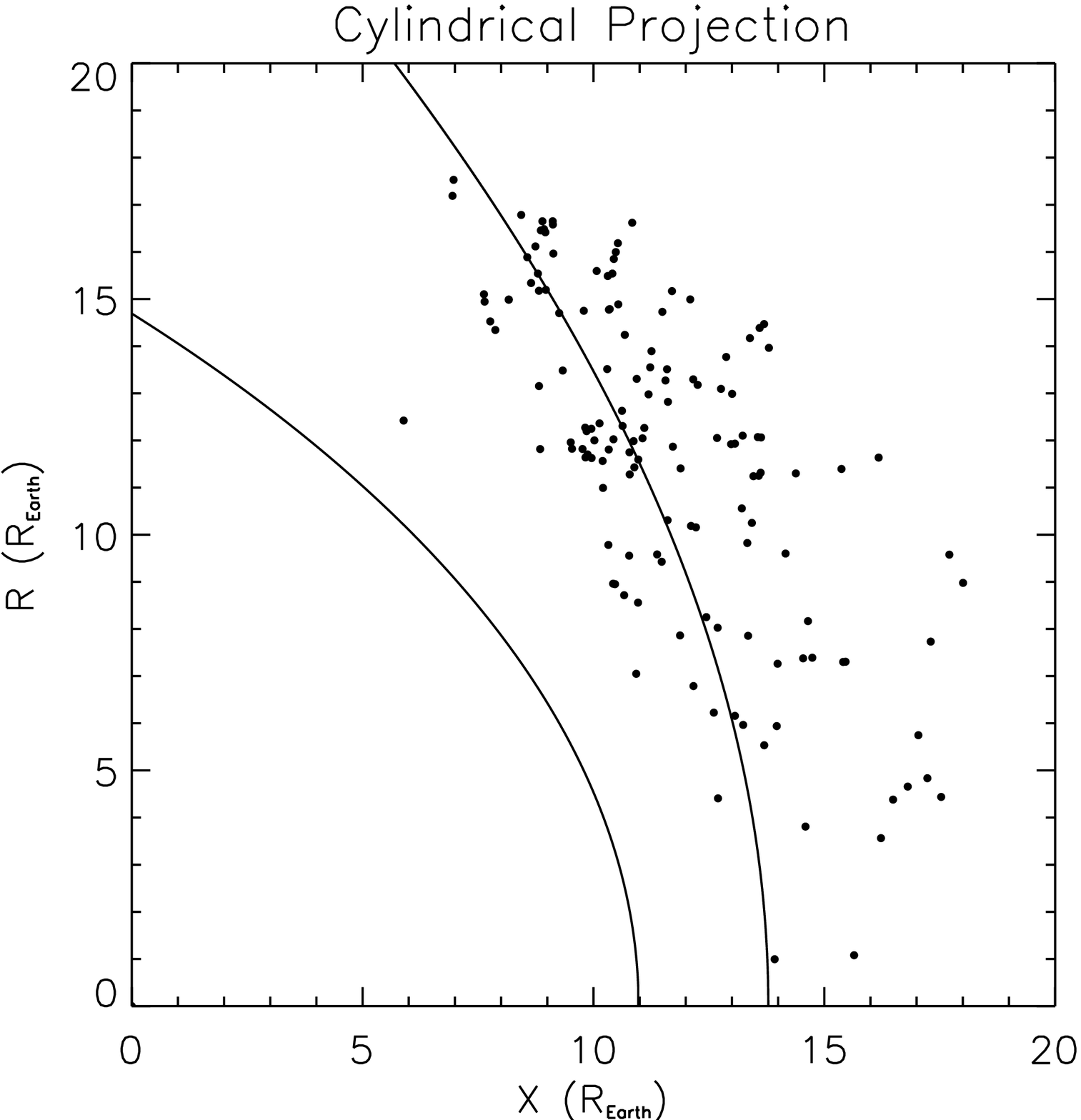,width=200pt}  
\caption{Cylindrical projection of Cluster spacecraft center positions during HFA observation and the average bow shock and magnetopause positions in GSE system. The shape of the magnetopause and the bow shock were calculated using the average solar wind pressure \citep{sibeck91:_solar, tsyganenko95:_model, peredo95:_three_alfven_mach}. The coordinates were plotted in $R_{\mathrm{Earth}}$ units. \citep[Based on][]{facsko09:_global_study_of_hot_flow}}
\label{fig:hfacl}
\end{figure}

\begin{figure*}[!t]
\centering
\begin{tabular}[t]{rl}
\multicolumn{2}{c}{\epsfig{file=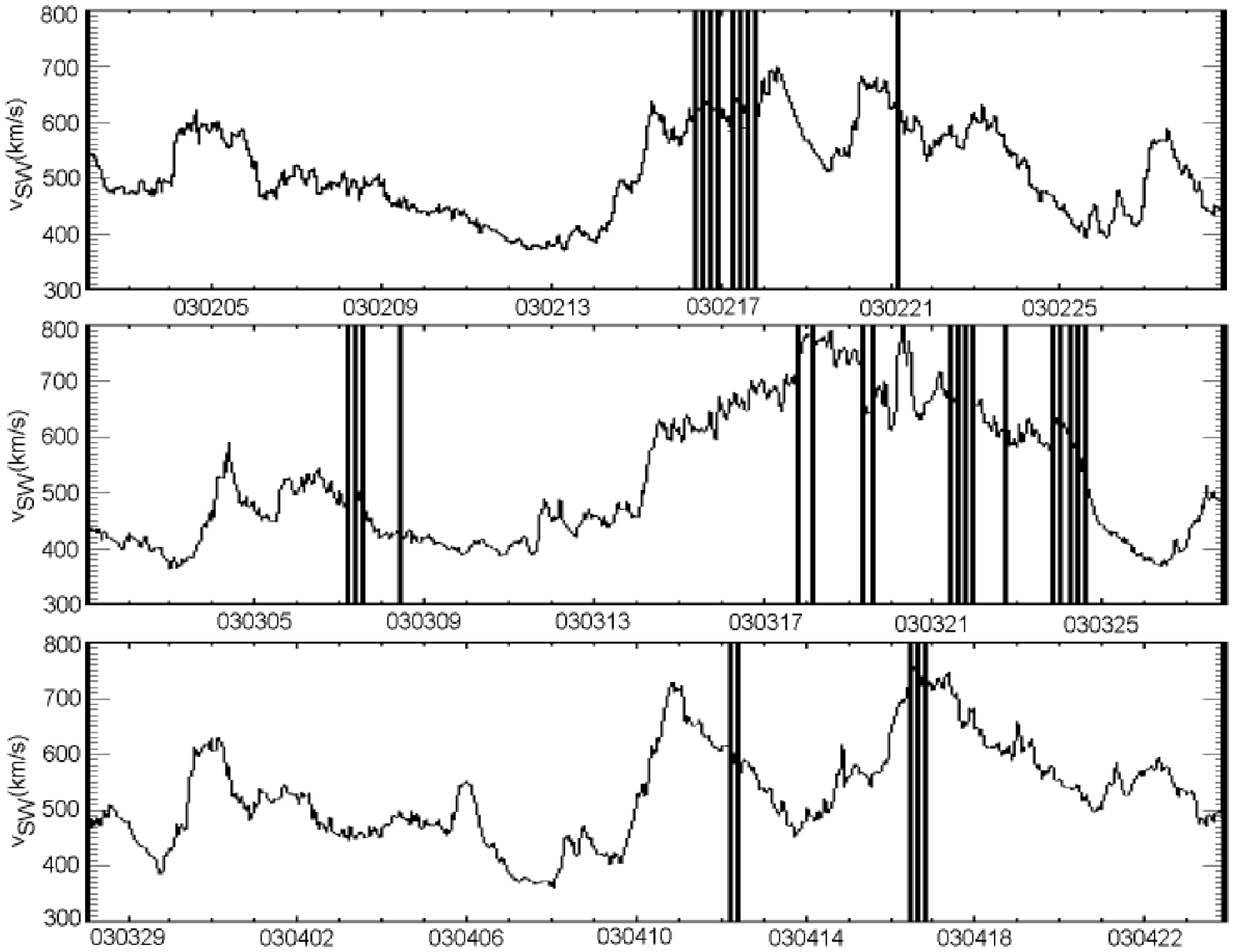,width=300pt}} \\
\epsfig{file=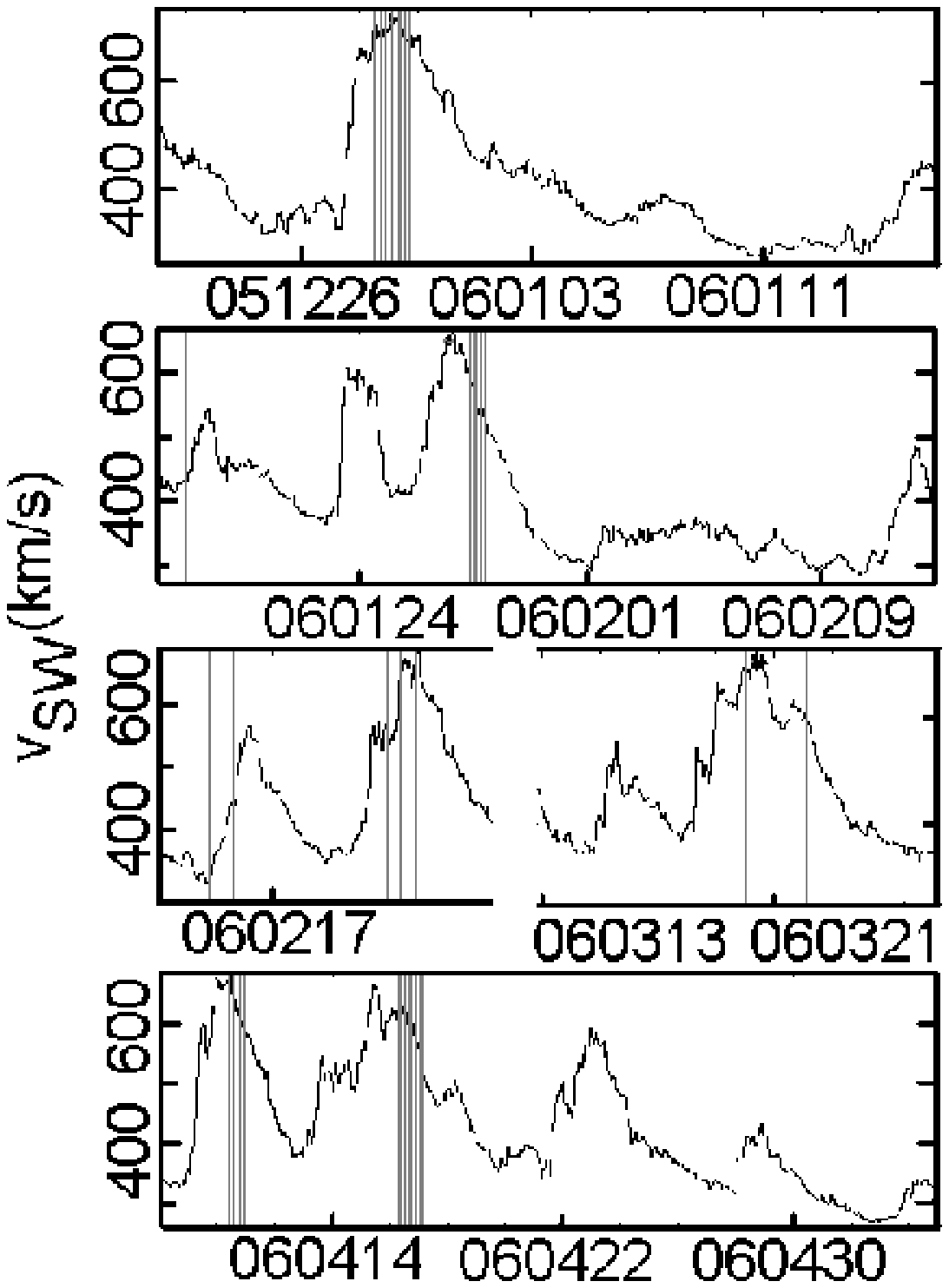,width=200pt} &
\epsfig{file=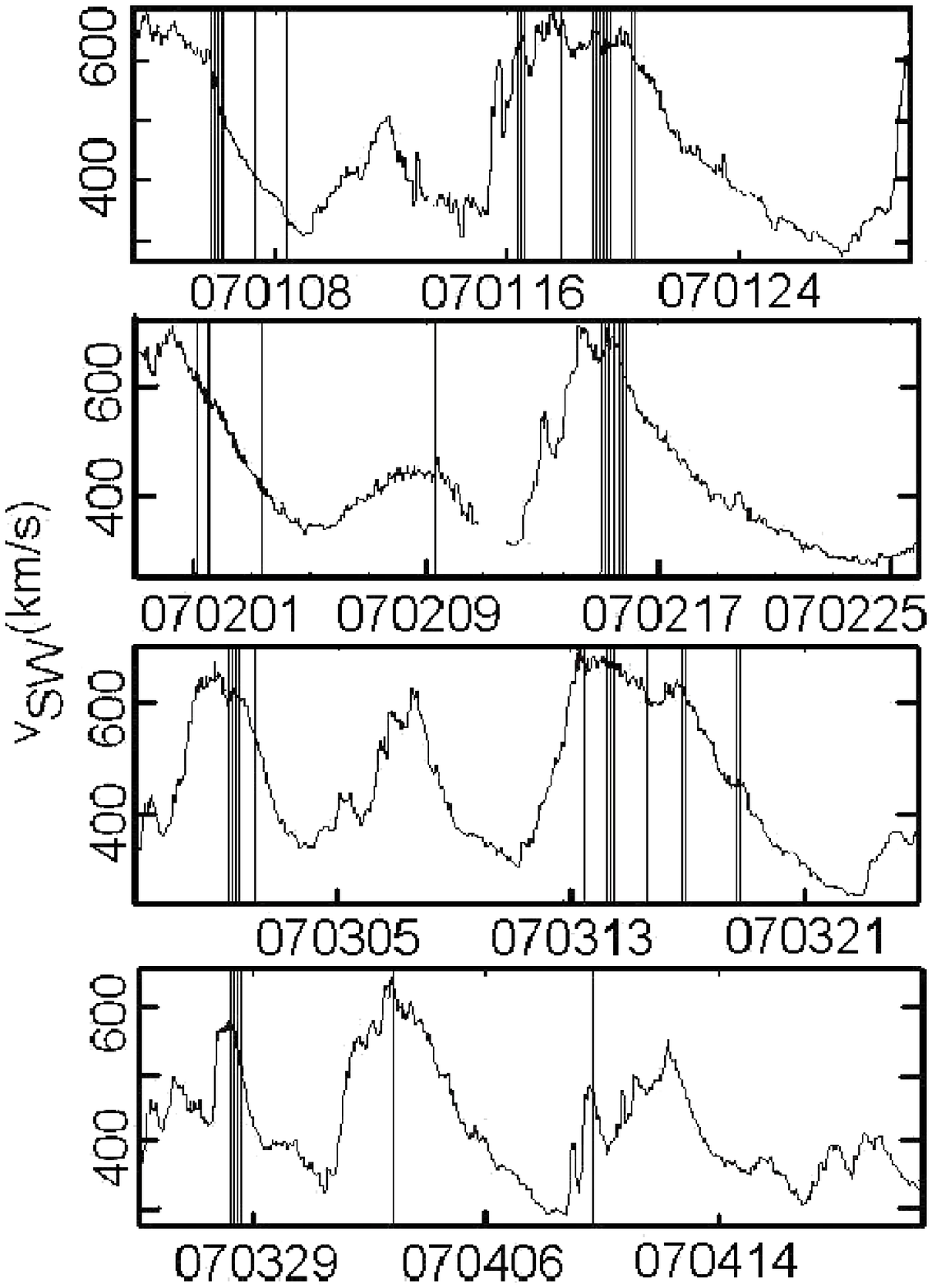,width=200pt} \\
\end{tabular}
\caption{1 hour averaged solar wind speed; the vertical red lines give the time of occurrence of HFAs. The top, bottom left and right figures were made using ACE SWEPAM instrument in 2003, 2006 and 2007, respectively. The connection between the fast solar wind regions and the HFAs is evident \citep{facsko09:_global_study_of_hot_flow}.}
\label{fig:vSWvel}
\end{figure*}

\begin{figure*}[!tb]
\centering
\epsfig{file=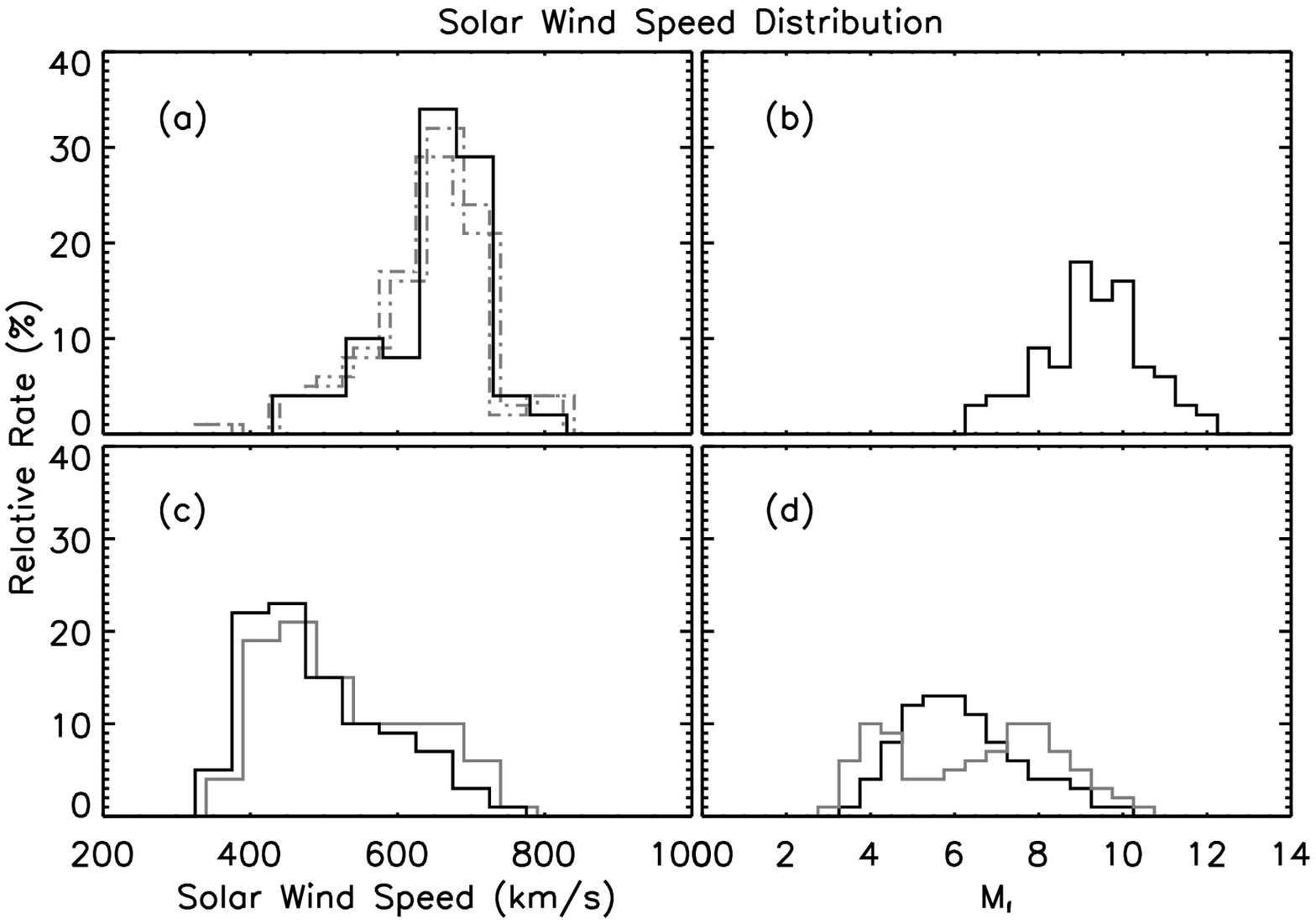,width=400pt}  
\caption{Upstream solar wind speed distribution measured by Cluster and ACE spacecraft. (a) Solar wind speed distribution measured by Cluster-1 (shifted grey dash dotted line) and Cluster-3 CIS HIA (shifted grey dash dotted line) ACE SWEPAM (solid line) upstream of HFA formation. (b) Fast magnetosonic Mach number distribution calculated using ACE MAG and SWEPAM data during HFA formation. (c) Solar wind speed distribution measured by ACE SWEPAM in 2003, 2006 and 2007, Spring (grey line) and from 1998 to 2008 (solid line). (d) Same as (c) but measured in $M_f$ units.}
\label{fig:vswdistr}
\end{figure*}

Using the criteria listed above 124 HFA events were found: between 2003 February and April 33 events, from 2005 December to 2006 April  41 events and finally between 2007 January and April another 50 events were identified (Fig.~\ref{fig:hfacl}). After plotting the ACE SWEPAM \citep[Solar Wind Electron, Proton, and Alpha Monitor;][]{mccomas98:_solar_wind_elect_proton_alpha} 1-hour averaged solar wind speed measurements and indicating the HFA events it seems that the HFAs appear when the solar wind speed is higher than the average (Fig.~\ref{fig:vSWvel}). This assumption is confirmed by the distribution of the solar wind speed and fast-magnetosonic Mach-numbers calculated using ACE MAG \citep[MAGnetometer;][]{smith98:_ace_magnet_field_exper} and SWEPAM measurements (Fig.~\ref{fig:vswdistr}). The solar wind speed is approximately 130\,km/s higher than the average during the HFA formation phase and the fast-magnetosonic Mach number is also higher than average. The difference is between 2.7-3.7\,$M_{\mathrm{f}}$ depending the year \citep{facsko09:_global_study_of_hot_flow}. This condition was also observed by \citet{safrankova00:_magnet} but they observed HFAs in the magnetosheath where the differences are not as significant as in the solar wind. Based on measurements made in 2003 it seems that a higher solar wind pressure can be also a condition of HFA formation. Finally after analyzing the events of two additional years (2006 and 2007) the solar wind pressure seems to be not a condition for HFA formation; furthermore it is lower then the average however the average particle density is similar \citep{facsko09:_global_study_of_hot_flow}. 

\subsection{The results of the statistical study}
\label{sec:results}

\begin{figure}[htb]
\centering
\epsfig{file=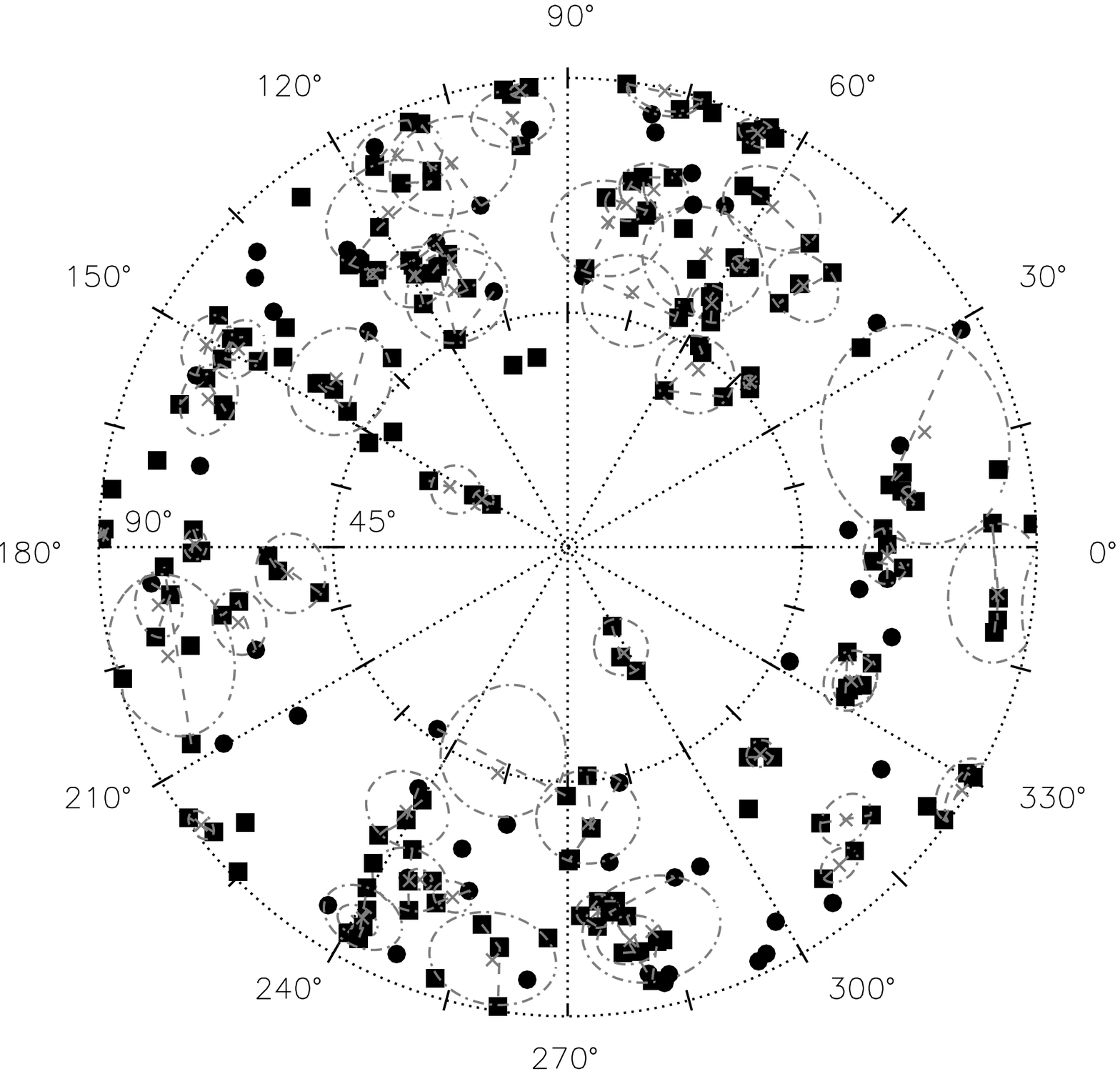,width=200pt}
\caption{Polar plot of the direction of the normal vectors of TDs. The azimuthal angle is measured between the GSE y direction and the projection of the normal vector onto the GSE yz plane. The distance from the center is the $\gamma$ angle as determined by the cross-product method. The TD normal vector is in a special polar coordinate system in which we measure the $\gamma$ angle from the center, and where the azimuth is the angle of GSE y and the projection of normal vector to GSE yz plane. The regions surrounded by dashed lines are the projection of error cones around the average normal vector marked by ``X''. Circles and squares symbolize ACE and Cluster data, respectively. \citep[Based on][]{facsko09:_global_study_of_hot_flow}}
\label{fig:anglesc}
\end{figure}

The most important result of the statistical study using Cluster multi-spacecraft is the discovery of the higher solar wind speed as a condition of HFA formation (Sec.~\ref{sec:selection}). This result is unexpected because the original purpose of the statistical study was to compare with \citet{lin02:_global}'s results. She performed global hybrid simulation of HFAs and presented size estimation, angle estimations and size-angle functions (See: Sec.~\ref{sec:theory}). The predicted size was confirmed by the study ($1-2\,R_{\mathrm{Earth}}$), the average value of $\Delta\Phi$ change angle was $~70^o$, the $\gamma$ angle between the solar direction and the TD normal was larger than $45^o$ as predicted (See Fig.~\ref{fig:anglesc}). This large gap around the Sun-Earth direction has been observed by \citet{schwartz00:_condit} before. Finally, the previously described size-angle dependencies were also confirmed. The results by \citet{schwartz00:_condit} on the condition of HFA formation was also confirmed because several part of their result was used for the size estimation. \citep{facsko09:_global_study_of_hot_flow} 

\section{Discussion}
\label{sec:discussion}

\begin{figure}[ht]
\centering
\epsfig{file=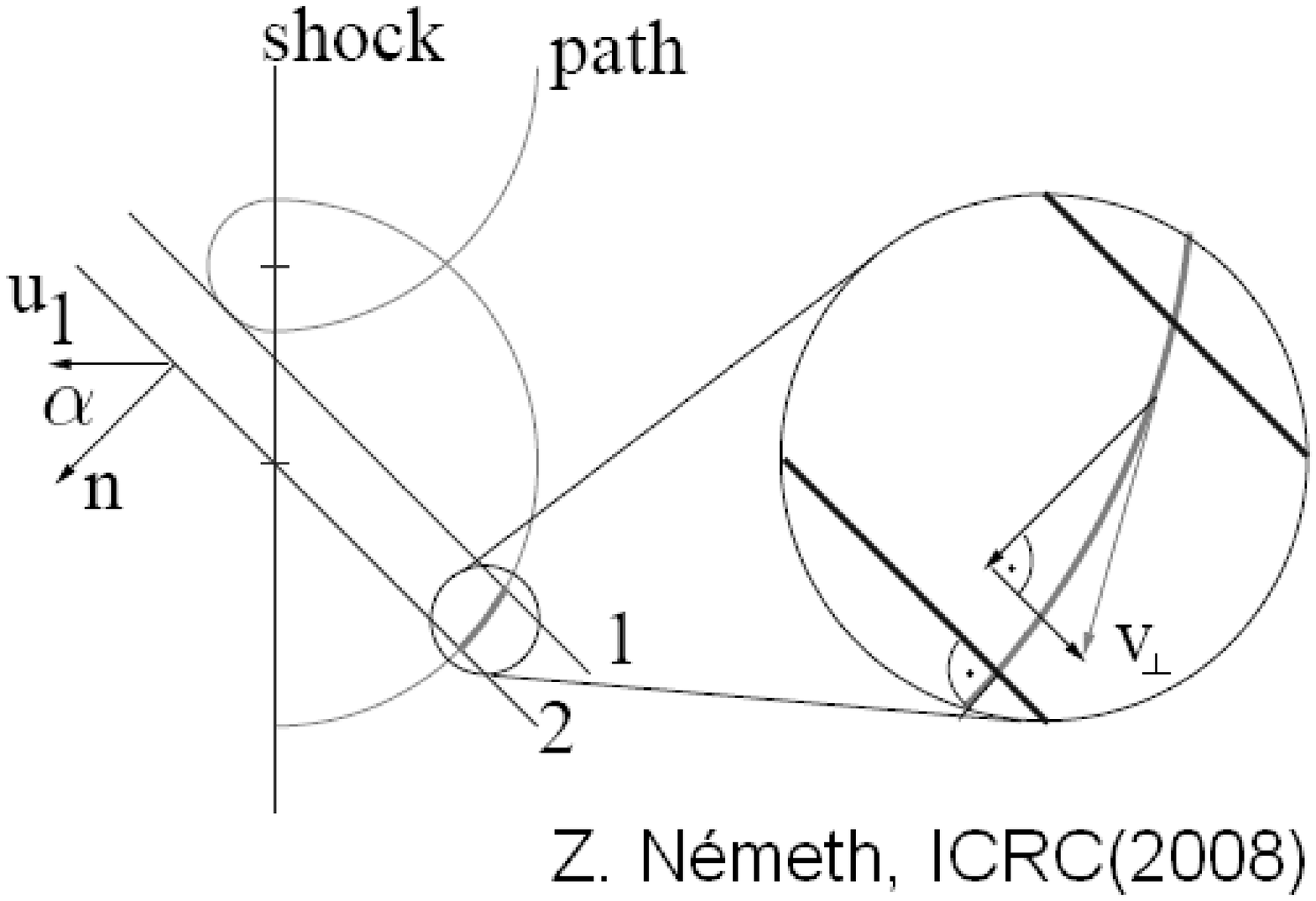,width=250pt}
\caption{The configuration of the bow shock and the tangentional discontinuity during a HFA event. The orbit of a trapped ion is shown. \citep[From][]{nemeth08:_partic_accel_at_inter_of}}
\label{fig:zoli}
\end{figure}

\citet{nemeth08:_partic_accel_at_inter_of} developed the following empirical criterion for the escape of energetic particle, after analyzing the geometry of escaping particles:
\begin{equation}
\label{eq:critical}
\sin\alpha\ge\frac{1}{\frac{B_1}{B_2}\left(1-\frac{u_1}{v_\perp}\right)},  
\end{equation}
where $\alpha$ is the angle of solar wind velocity and discontinuity normal, $1$ and $2$ are two possible positions of the discontinuity, $B_1$ and $B_2$ are the magnetic field at these positions, $u$ is the solar wind velocity and $v_\perp$ is the component of the particle velocity which is perpendicular to the magnetic field. If the solar wind speed is considered infinity and the rate of the magnetic field magnitude in the upstream and downstream region is set to the typical value (4) then this angle is $\alpha=41.8^o$ which agrees well with simulation results \citep{lin02:_global} and observations \citep{facsko08:_statis_study_of_hot_flow,facsko09:_global_study_of_hot_flow,facsko09:_clust_hot_flow_anomal_obser,schwartz00:_condit}. Unfortunately this formula does not explain why the observed speed is necessary for HFA formation. Further test particle and 3D hybrid simulations are needed to explain this condition theoretically.

\section{Summary}
\label{sec:summary}

The Cluster spacecraft have allowed us to observe HFAs with unprecedented resolution, providing a wealth of information and allowing us to compare the observations with simulations. Using a sample of 124 HFAs observed by Cluster we get the following general features:
\begin{enumerate}
\item The cone angle of the TD normal are high and the large gap around the Sun-Earth direction in the TD normal has no anisotropy.
\item {\label{res2}} Both the higher solar wind speed and Mach number are conditions of HFA formation. The velocity increase seems to be less than 200km/s (measured in 2003) or 130km/s (measured in 2006 and 2007). The relative increase is larger when measured in fast magnetosonic Mach number units. 
\item We estimated the size and got good agreement with the simulations of 1\,$R_\mathrm{Earth}$.
\item We constructed size-TD normal cone angle, Size-B direction change angle and Size-solar wind Mach number plots. All of them confirm the result of hybrid simulations.
\item Our theory presented in Sec.~\ref{sec:discussion} develops the higher solar wind condition from fundamental geometrical features of HFA formation. 
\end{enumerate}
Further research is needed to explain more satisfactory the reason for condition (\ref{res2}). 3D hybrid simulations might give better insight here. 

Currently HFAs are receiving widen interest again. We suggest some unsolved problem in the field of HFAs:
\begin{enumerate}
\item The Kronian HFA analogies do not agree with the geometric features based on terrestrial events. It would be useful to detect more events and perform a statistical study. Furthermore it is very disturbing that Kronian analogies exhibit an increase, rather than a decrease, in the particle density. The magnetic field signature however, is similar to that at Earth. Unfortunately it remained unclear what TD-bow shock interaction produces in a different region of parameter space; for example whether the much weaker electric field at Saturn leads to a different type of phenomenon. This problem and the fact that only HFA-like events have been reported to date, also might motivate further studies of HFAs at other planets. 
\item HFA analogies were discovered at the Mars, Venus, Saturn and probably also in the inner heliosheath. It would be nice to detect them at bow shock of a fast moving star or extragalaxy. The beam formed by the interaction of the discontinuity interacts with the interstellar or intergalactic matter and this might observed. It would be nice to construct a model for this. 
\item Since the launch of THEMIS, two multi-spacecraft fleets can observe HFAs. It would be most interesting to observe them simultaneously to study their temporal development. It is possible in 2007, 2008 and 2009. It would be interesting to observe the same event with two distant satellites: one (or more) might be one of the THEMIS or Cluster spacecraft and another may be Geotail or other satellite. 
\item Finally it would be useful to observe simultaneously a hot flow anomaly at different scales to understand better the turbulence inside the cavity. The proposed Cross Scale mission may perform this observation. 
\end{enumerate}
After a short break the research on HFAs seems these structures again give exciting results. In spite of numerous papers and results we are only the beginning of understanding these complex features. 

\section*{Acknowledgements}
\label{sec:acknow}

G\'a\-bor Facs\-k\'o acknowledges useful dis\-cus\-si\-ons with K.~Kecs\-ke\-m\'e\-ty, G.~Er\-d\H{o}s, M.~T\'atrallyay and Z.~N\'emeth. The author thanks the ACE MAG (PI: N.~F.~Ness) and SWEPAM (PI: D.~J.~McComas) working teams for the magnetic field and plasma data; as well as the help of Thierry Dudok de Wit and Anna-M\'aria V\'\i gh for improving the English of this paper. Fig.~\ref{fig:hfasc} was made with the QSAS science analysis system provided by the United Kingdom Cluster Science Centre (Imperial College London and Queen Mary, University of London) supported by The Science and Technology Facilities Council (STFC). This work was supported by the OTKA grant K75640 of the Hungarian Scientific Research Fund. 

\bibliography{asr-s-09-00015-tx}
\bibliographystyle{elsarticle-harv}

\end{document}